\documentclass[twocolumn,reprint, aip,amssymb,apl, superscriptaddress]{revtex4-1}
\usepackage{amsmath}
\usepackage{amssymb}
\usepackage{amsthm}
\usepackage{graphicx}

\begin{document}
\title{Ultimate fast optical switching of a planar microcavity in the telecom wavelength range}
\author{Georgios Ctistis}\email{g.ctistis@utwente.nl}
\affiliation{Complex Photonic Systems (COPS), MESA+ Institute for
Nanotechnology, University of Twente, 7500 AE Enschede, The Netherlands}
\author{Emre Yuce}
\affiliation{Complex Photonic Systems (COPS), MESA+ Institute for
Nanotechnology, University of Twente, 7500 AE Enschede, The Netherlands}
\affiliation{Center for Nanophotonics, FOM Institute for Atomic and
Molecular Physics (AMOLF), 1098 XG Amsterdam,
The Netherlands}
\author{Alex Hartsuiker}
\affiliation{Complex Photonic Systems (COPS), MESA+ Institute for
Nanotechnology, University of Twente, 7500 AE Enschede, The Netherlands}
\affiliation{Center for Nanophotonics, FOM Institute for Atomic and
Molecular Physics (AMOLF), 1098 XG Amsterdam,
The Netherlands}
\author{Julien Claudon}
\affiliation{CEA-CNRS-UJF "Nanophysics and Semiconductors" joint
laboratory, CEA/INAC/SP2M, 38054 Grenoble
Cedex 9 France}
\author{Maela Bazin}
\affiliation{CEA-CNRS-UJF "Nanophysics and Semiconductors" joint
laboratory, CEA/INAC/SP2M, 38054 Grenoble
Cedex 9 France}
\author{Jean-Michel G\'erard}
\affiliation{CEA-CNRS-UJF "Nanophysics and Semiconductors" joint
laboratory, CEA/INAC/SP2M, 38054 Grenoble
Cedex 9 France}
\author{Willem L. Vos} \email{W.L.Vos@tnw.utwente.nl}
\affiliation{Complex Photonic Systems (COPS), MESA+ Institute for
Nanotechnology, University of Twente, 7500 AE Enschede, The Netherlands}

\begin{abstract}
We have studied a GaAs--AlAs planar microcavity with a resonance near 1300 nm in the telecom range by ultrafast pump-probe reflectivity. By the judicious choice of pump frequency, we observe a ultimate fast and reversible decrease of the resonance frequency by more than half a linewidth due to the instantaneous electronic Kerr effect.
The switch-on and switch-off of the cavity is only limited by the cavity storage time of $\tau_{cav}=0.3 ps$ and not by intrinsic material parameters. Our results pave the way to supra-THz switching rates for on-chip data modulation and real-time cavity quantum electrodynamics.
\end{abstract}

\maketitle
%

Switches are widely applied and necessary ingredients in modulation and computing schemes~\cite{Sze1985}.
The recent progress on photonic integrated circuits~\cite{Noda2003, Assefa2010} promises to overtake boundaries set by conventional switching technology.
To do so, ultrafast switching of photonic cavities is crucial as it allows the capture or release on demand of photons~\cite{Johnson2002ab, Almeida2004a, Tanabe2007}, which is relevant to on-chip communication with light as information carrier~\cite{Yanik2004ab}, and to high-speed miniature lasers~\cite{Hense1997aa}.
Ultrafast switching would also permit the quantum electrodynamical manipulation of coupled cavity-emitter systems~\cite{Gerard2003} in real-time.
Switching the optical properties of photonic nanostructures is achieved by changing the refractive index of the constituent materials.
To date, however, the switching speed has been limited by material properties~\cite{Leonard2002aa, Mazurenko2005aa, Harding2007aa, Hu2008aa}, but not by optical considerations.
To achieve ultimate fast switching of a cavity two challenges arise. Firstly, both the switch-on
and switch-off times $\tau_{on}$ and $\tau_{off}$ must be shorter
than all other relevant time scales for the system, i.e., the cavity
storage time in photon capture/release experiments $\tau_{cav}$, 
or the vacuum Rabi period $\tau_{Rabi}$ for a strongly
coupled emitter-cavity system \cite{Khitrova2006aa}. Secondly, the
refractive index change must be large enough to switch the cavity
resonance by at least half a linewidth.

Here, we demonstrate the ultimate fast switching of the resonance of a planar cavity in the well-known GaAs/AlAs system in the telecom wavelength range.
We exploit the instantaneously fast electronic Kerr effect by the judicious tuning of the pump and probe frequencies relative to the semiconductor bandgap.
We observe that the speed of the switching is then only limited by the dynamics of the light in our cavity ($\tau_{cav}$ = 0.3 ps), but not by the intrinsic material parameters.

\begin{figure}[htb]
\begin{center}
 \includegraphics[]{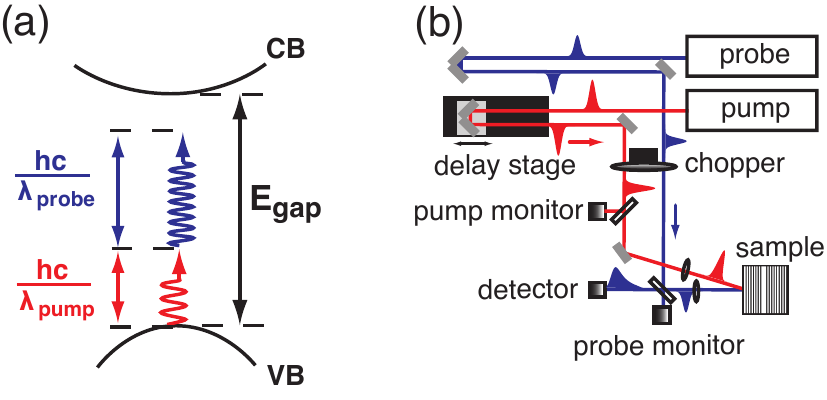}
   \caption{ (Color online) (a) Schematic energy diagram. The frequency of the probe beam is resonant with the cavity. The pump frequency is tuned such that the sum of pump and probe energies are less than the energy gap of GaAs to avoid two-photon absorption. (b) Schematic of the set-up. The probe-beam path is shown in blue, the pump-beam path in red.
 }
 \label{Fig1}
 \end{center}
 \end{figure}

Instantaneous on- and off-switching with
vanishing $\tau_{on}$ and $\tau_{off}$ is feasible with the
well-known nonlinear refractive index from nonlinear optics
\cite{Boyd1992aa}. Physically the electronic Kerr effect is the
fastest Kerr phenomenon on account of the small electron mass.
In many practical situations, however, non-degenerate
two-photon absorption overwhelms any instantaneous effect and
therefore also the dispersive electronic Kerr effect
\cite{Boyd1992aa,Harding2007aa}.
In order to avoid two-photon absorption and to access the electronic Kerr
switching regime, we designed our experiment to operate with low energy pump
photons, see Fig.~\ref{Fig1}~(a).
First of all, the energy of the pump photons is chosen below half the semiconductor
band gap energy ($E_{pump}<\frac{1}{2}E_{gap}$). Secondly, the energies of the pump and probe
photons are chosen so that their sum does not exceed the bandgap energy
of the semiconductor ($E_{probe}+E_{pump}\leq E_{gap}$) \cite{Hartsuiker2008ab}.
These conditions serve to suppress instantaneous two-photon
generation of free carriers and to perform electronic Kerr switching at elevated frequencies
($E_{probe}>\frac{1}{2}E_{gap}$), including the telecom band, with a broad range of semiconductors.

We have performed our experiments on a planar microcavity grown by means of molecular-beam epitaxy. 
The sample consists of a $\lambda$-thick GaAs layer ($\rm{d=376\ nm}$) sandwiched between two Bragg stacks consisting of 7 and 19 pairs of $\lambda/4$-thick layers of nominally pure GaAs ($\rm{d_{GaAs}=94\ nm}$) or AlAs ($\rm{d_{AlAs}= 110\ nm}$). 
The cavity was designed such that the resonance occurs at $\lambda_0$ = 1280 nm in the Original (\emph{O}) telecom band. 
Measuring the linewidth of the cavity resonance we obtained a quality factor $\rm{Q}=320 \pm 30$ corresponding to a cavity storage time of $\tau_{cav}=0.3\ \rm{ps}$.

To perform Kerr-switching on a semiconductor microcavity, we have built a set-up illustrated in Fig~\ref{Fig1}~(b), consisting of two independently tunable optical parametric amplifiers (OPA,Topas) that are the sources of the pump and probe beams. 
The pump beam can be tuned down to 4200 $\rm{cm^{-1}}$ (2400 nm) and the probe beam is tuned to cavity resonance at 7810 $\rm{cm^{-1}}$ (1280 nm). 
Under these conditions, we solely pump GaAs (the refractive index change of AlAs is much smaller and can be neglected here).
The pump beam has a larger Gaussian focus (75 $\mu$m full width at half maximum) than the probe beam (28 $\mu$m), to ensure that only the central flat part of the pump focus is probed and that the probed region is homogeneously pumped.
The OPAs have pulse durations $\tau_{\rm{P}} = 0.12 \pm 0.01$ ps. The delay $\rm{\Delta t}$ between pump and probe pulse is set by a delay stage with a resolution of 0.01 ps.
A versatile measurement scheme was developed to compensate for possible pulse-to-pulse variations in the output of our laser \cite{Euser2009aa}.

Figure 2 shows transient reflectivity versus frequency spectra for three different pump-probe delay times. 
One can see that the cavity resonance red-shifts when approaching pump-probe overlap. Afterwards, at positive delay times, the cavity resonance blue-shifts back to its original frequency. 
By modeling the transient reflectivity trough with a Lorentzian, both the resonance frequency $\omega_0$ and the width of the resonance $\Gamma$ are obtained for every time delay $\Delta t$.
The unswitched cavity resonance frequency ($\omega_0=7810\ cm^{-1}$) and width ($\Gamma=12\ cm^{-1}$)
are obtained at a delay of $\Delta t= -5\ \textrm{ps}$, confirming a cavity quality factor of $\textrm{Q} = 320$. 
 
\begin{figure}[htb]
\begin{center}
\includegraphics{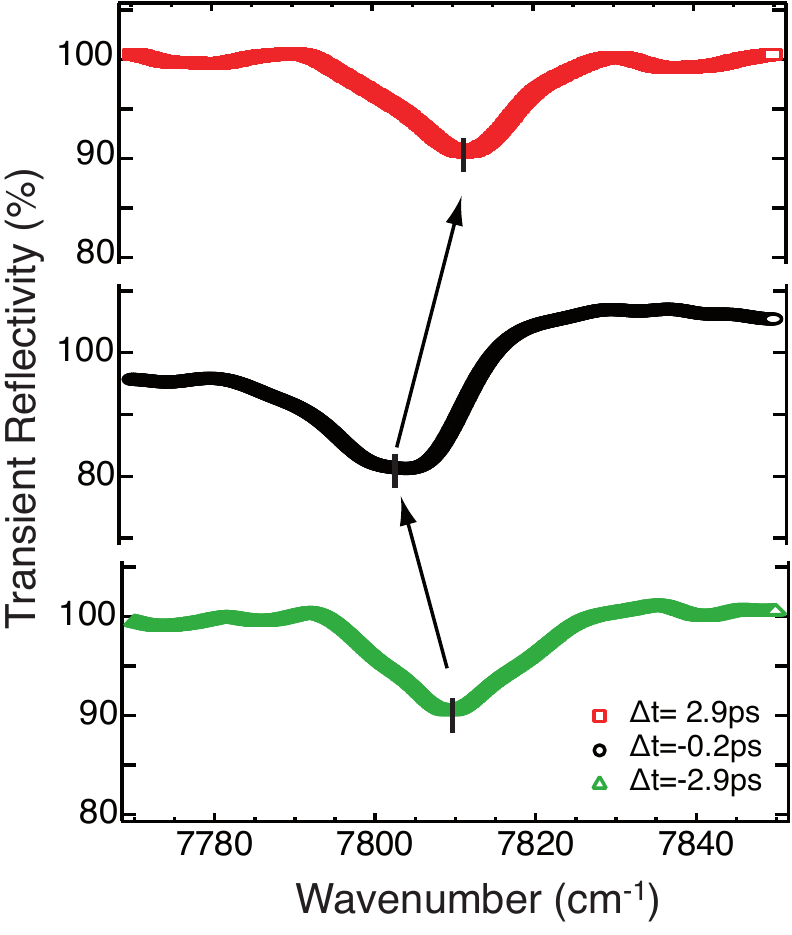}
\caption{(Color online) Transient reflectivity versus frequency spectra for three different delay times, $\Delta t=-2.9\ ps$ (triangles), $\Delta t=-0.2\ ps$ (circles), and $\Delta t=+2.9\ ps$ (squares). 
We observe an ultrafast shift of the cavity resonance to lower frequencies and back to its original frequency.
}
\label{Fig2}
\end{center}
\end{figure}

\begin{figure}[htb]
\begin{center}
\includegraphics{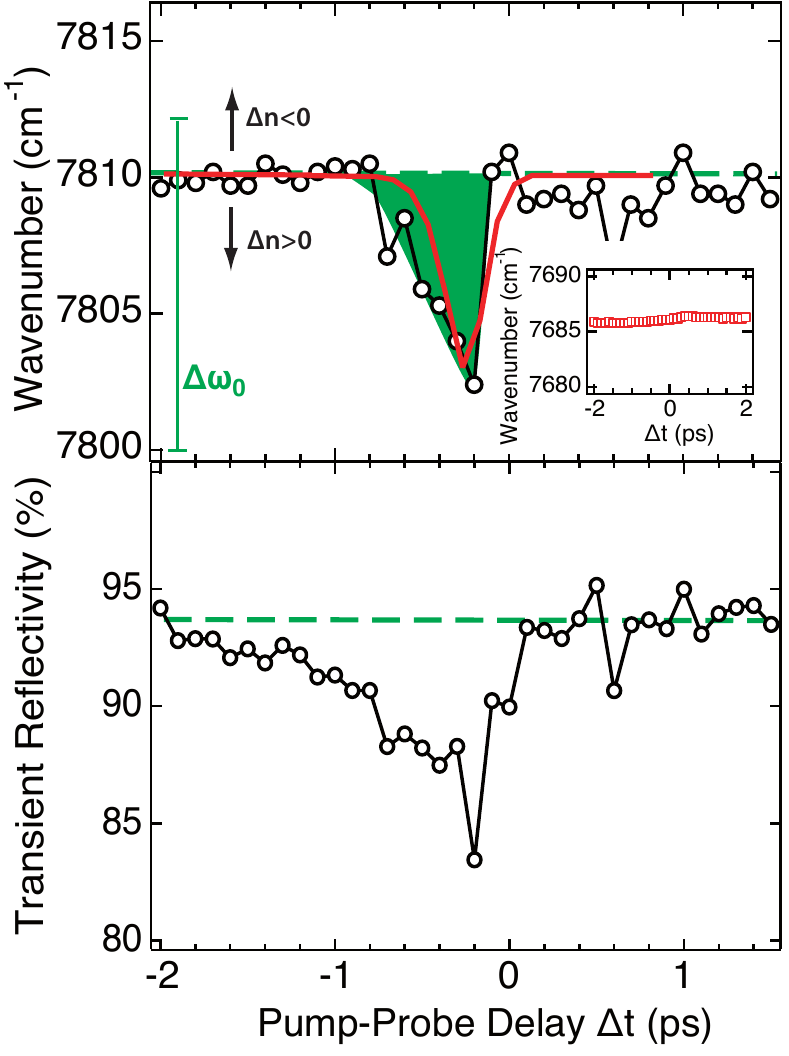}
\caption{(Color online) (upper panel) Measured (open symbols) and modeled (line) resonance frequency vs. pump-probe delay. The pump frequency is at $4166\ cm^{-1}$. 
The resonance frequency reaches its maximum shift at -0.2 ps and quickly returns to its unswitched value. The model matches the experiment well. 
The inset shows no switch when the pump frequency is at $5000\ cm^{-1}$ for comparison.  
(lower panel) Transient reflectivity vs. pump-probe delay at the cold cavity resonance frequency. 
The transient reflectivity shows similar reversible behavior as the resonance. 
}
\label{Fig3}
\end{center}
\end{figure}

Figure~\ref{Fig3}~(a) shows the switching of the cavity resonance due to the electronic Kerr effect. 
The resonance frequency as a function of delay near pump and probe
coincidence is obtained from spectra similar to those shown in Fig~\ref{Fig2}. 
The dashed horizontal line denotes the unswitched resonance frequency $\omega_0$. 
The measured resonance frequency shifts to a lower frequency starting at -0.8 ps and reaching its maximum shift at -0.2 ps.
Subsequently, the resonance frequency shifts back to its original frequency $\omega_0$.
The dynamic frequency shift $\Delta\omega=7\ cm^{-1}$ clearly exceeds one half of the cold cavity linewidth.
We conclude from the shift to a lower frequency that the refractive index is increased by $0.1\ \%$, which corresponds to a positive Kerr coefficient for GaAs \cite{Hartsuiker2008ab}. 
To confirm our interpretation we performed calculations with a dynamic Fabry-P\'{e}rot model
 \cite{Marzenell2000aa,Born2002}, taking solely into account the refractive index change of GaAs due to the electronic Kerr effect. We observe an excellent quantitative agreement between the measured and
calculated shifts for the amplitude and the temporal evolution.
Our results furthermore demonstrate that our method serves to truly inhibit excitation of free carriers since there is no blue-shift of the resonance at $\Delta t > 0\ ps$. 
For comparison, the inset shows the result when the pump frequency is increased to $5000\ cm^{-1}$ in the free carrier regime.
There is no instantaneous shift of the resonance in the given range as expected; the free-carrier excitation leads to a blue-shift after a time of $3\ \rm{ps}$ \cite{Harding2007aa}.
The very good agreement between the calculation and our experimental results firmly confirm the ultimate fast electronic Kerr switching of our photonic microcavity.

Figure \ref{Fig3}(b) shows the transient reflectivity  of the unswitched resonance frequency as a function of pump-probe delay.
The relatively high reflectivity of the trough ($R_{trough}=90\ \%$) is a result of the asymmetric cavity design. 
One can see that the transient reflectivity quickly changes from high to low (at pump-probe overlap) and back to high, within $1.5\ ps$.
This decrease is a result of the change of the refractive index of GaAs not only in the cavity but also in the mirrors, which leads to a higher contrast in the Bragg stack.

The shape of the resonance shift in Fig.~\ref{Fig3}~(a) is strikingly asymmetric. The asymmetry is a direct result of the light dynamics in the cavity.
In our experiment we are probing the change in the dielectric function $\Delta\varepsilon$
of the material induced by $\textbf{P}^{(3)}$ \cite{Boyd1992aa}. 
The magnitude of the observed frequency shift $\Delta\omega$ is given by the overlap integral of the pump and probe,
the latter has been stored in the cavity and subsequently escaped, at a certain delay time $\Delta t$:
\begin{eqnarray}
\frac{\Delta\omega (\Delta t)}{\omega_0(0)} = \frac{ \Delta\varepsilon (\Delta t)}{\varepsilon (0)}
= \frac{\chi^{(3)}}{\varepsilon (0)} \int{E(\omega_0,t^{\prime}-\Delta t) E(\omega_2,t^{\prime})d t^{\prime}}
\label{eq:overlap}
\end{eqnarray}

\noindent Hence, we tomographically sample the probe with the shorter pump. Thus, while scanning $\Delta t$, we obtain a shift that follows the cavity envelope. 
The shift is maximal at $-0.2\ ps$ when the cavity field is also maximum.

We have demonstrated for the first time the switching of a
semiconductor microcavity at telecom wavelengths using the
electronic Kerr effect. 
Our system also serves as a model system, since the nature of the switch process can be employed in any realization of a semiconductor cavity.
We observed that the switching speed is limited by the cavity
storage time and not by material properties. The 0.3 ps storage time in our work
paves the way to sub-ps real-time data modulation. Since
switching using the electronic Kerr effect is repeatable, on-chip supra-THz switching rates are feasible. 
Our results also open an avenue towards ultrafast control of all solid-state cavity quantum electrodynamical systems that exploit the strong coupling regime of quantum wells or quantum dots in optical microcavities \cite{Khitrova2006aa}.

We thank Allard Mosk, Harm Dorren, Huib Bakker, and Pepijn Pinkse for stimulating
discussions. This research was supported by Smartmix Memphis, NWO-Vici
(to WLV), and the QSWITCH ANR project (to JMG). This work is also part of the research
program of FOM, which is financially supported by NWO.

\end{document}